\documentclass[submission,copyright,creativecommons]{eptcs}
\providecommand{\event}{ACL2 2018} 
\usepackage{breakurl}             
\usepackage{underscore}           
\usepackage{comment}
\usepackage{fancyvrb}

\usepackage{float}
\floatstyle{boxed}
\restylefloat{figure}

\title{A Toolbox For Property Checking From Simulation Using Incremental SAT (Extended Abstract)}
\author{Rob Sumners
\institute{Centaur Technology}
\email{rsumners@centtech.com}
}
\def\titlerunning{Property Checking From Simulation Using Incremental SAT}
\def\authorrunning{Rob Sumners}
\begin{document}
\maketitle

\begin{abstract}
  We present a tool that primarily supports the ability to check bounded
  properties starting from a sequence of states in a run. The target design is
  compiled into an AIGNET which is then selectively and iteratively translated
  into an incremental SAT instance in which clauses are added for new terms and
  simplified by the assignment of existing literals. Additional applications of
  the tool can be derived by the user providing alternative attachments of
  constrained functions which guide the iterations and SAT checks
  performed. Some Verilog RTL examples are included for reference.
\end{abstract}

\section{Overview} \label{sec:overview}


Formal property verification of modern hardware systems is often too onerous to
prove directly with a theorem prover and too expensive to prove with more
automatic techniques. A common remedy to this difficulty is to focus on proving
properties from a more limited logic (e.g. STE~\cite{ste}, BMC~\cite{bmc},
using SVTVs~\cite{svtv} in previous ACL2 work). In these cases, properties are
reduced in scope to only consider the effects of a bounded number of next-steps
of the transition function of the hardware design; this, in turn, greatly
increases the capacity of the hardware designs one can target with automatic
propositional logic verification techniques (primarily SAT). The general
downside of this reduction in property scope is the need to codify and prove
(inductive) invariants of the reachable states of the design. The definition
and proof of these invariants is not only time consuming but the invariants are
often brittle and require significant maintenance with ongoing design
change. An alternative approach is to bypass the definition and proof of
invariants and instead check the properties from a set of representative states
of the system -- one can then either prove that these states are in fact
representative or accept the results of the property checks as a sufficient
semi-formal check of the design and disregard a full proof. We briefly cover
the design and definition of a tool (named {\tt exsim}) which builds upon
existing hardware verification work in the ACL2 community (VL~\cite{svtv},
SV~\cite{svtv}, AIGNET~\cite{aignet}) to add the capability to efficiently
check bounded properties across a sequence of next-states of the design using
incremental SAT~\cite{incremental}. An initial version of the tool and some
example Verilog designs and runs are included in the supporting materials.

\begin{figure}
\scriptsize
\begin{verbatim}
module toy (input            reset,             module top (input       reset,
            input            clk,                           input       clk,
            input            op,                            input       wave_op,
            input [1:0]      in,                            input [1:0] free_in,
            output reg [1:0] out);                          output      fail_out);
   ...                                              ...                              
  always@* w1 = tmp|in;                            always@(posedge clk) in <= free_in;
  always@* w2 = tmp&{2{in[0]}};                    always@(posedge clk) op <= wave_op;
  always@(posedge clk) tmp <= in;                  toy des(.*);
  always@(posedge clk) out <= (op ? w1 : w2);      always@(posedge clk) fail_out <= &out;
endmodule // toy                                endmodule // top  
\end{verbatim}
\normalsize
\caption{Definition of a simple {\tt toy} module and {\tt top} environment}
\label{fig:toy}
\end{figure}

\section{Design and Toy Example} \label{sec:design}

The {\tt exsim} function is split into two steps. The first step is the
function {\tt exsim-prep} which reads in a verilog design and compiles it into an
AIGNET -- an optimized ``circuit'' representation built from {\tt AND} and {\tt
  XOR} gates and inverters supporting efficient simplification
procedures~\cite{aignet}.

The second step of {\tt exsim} is the function {\tt exsim-run} which reads in a
waveform for the design (from a VCD file) and initializes the data structures
needed to iterate through the {\tt exsim-main-loop}. As part of this
initialization, inputs of the design are tagged as either {\tt :wave}, {\tt
  :rand}, or {\tt :free} corresponding to input values tracking values from the
input waveform, being generated randomly, or being left as free variables. In
addition, certain signals are tagged as {\tt :fail} which merely designates
these signals as the properties we are checking -- namely that these fail
signals never have a non-zero value after reset. A simple toy example in
Figure~\ref{fig:toy} is provided for reference -- the default signal tagging
function would tag {\tt free\_in} as the sole {\tt :free} input and {\tt
  fail\_out} as the sole {\tt :fail} signal to check.

The primary function of the {\tt exsim-main-loop} is to update an incremental
SAT instance~\cite{ipasir} such that the clauses in the SAT instance correspond
to a check of {\tt :fail} signals at certain clock cycles having the value of
$1$ relative to certain values of {\tt :free} inputs on previous clock
cycles. The main-loop iterates through a sequence of operations primarily
comprised of {\tt :step-fail}, {\tt :step-free}, and {\tt :check-fails} which
respectively add clauses to SAT for the :fail at the next clock cycle, add unit
clauses to SAT corresponding to an assignment to :free variables on an earlier
clock cycle, and calling the SAT solver to check if any of the :fail signals
could be $1$ in the current context. The ``compiled'' AIGNET from {\tt
  exsim-prep} allows for an efficient implementation of {\tt :step-free} and
{\tt :step-fail} for larger designs.

The choice of which step to take is heuristic and depends primarily on the
current number of clauses in the database. In particular, {\tt :step-fail} will
add clauses to SAT while {\tt :step-free} will reduce clauses in SAT and
heuristics will choose when to perform each action depending on how extensively
the user wishes to search for fails in a given instance. The default heuristic
choice function (as well as several other functions defining default operation)
can be overridden using ACL2's {\tt defattach} feature~\cite{defattach}.

Returning to our simple toy example from Figure~\ref{fig:toy}. Let's assume
that the input {\tt wave\_op} gets a value of $1$ at every clock cycle from
the input waves. After a sufficient number of {\tt :step-fail}s, this would
reduce the check of {\tt fail\_out} being $1$ to {\tt free\_in} having at least
a single $1$ in each bit position over $2$ consecutive clocks, $3$ clocks earlier.



The intent of the {\tt exsim} toolkit is to provide the user the capability to
control the SAT checks that are performed to optimize effective search capacity
over runtime. This is primarily achieved by allowing the user to carefully
specify which input variables should be :free (which reduces potential decision
space during SAT along with increasing propagations) and when to either extend
or reduce the current set of SAT clauses by inspecting the current number of
active clauses. The additional benefit of incremental SAT is that the aggregate
cost of successive searches is reduced by the sharing of learned reductions.

The {\tt exsim} toolkit is a work in progress. Current development efforts
afford better control over bindings of free variables and more intermediate
targets for search (not just {\tt :fail} signals). Related to these efforts, an
earlier proof of ``correctness'' needs to be reworked to keep apace with the
design changes. As a note on this ``proof'', the goal is to ensure that each
{\tt :check-fails} which returns UNSAT ensures no possible assignment for the
free variables in the scope of the current check. This reduces to proving an
invariant relating the current SAT clauses with the values of signals at
certain clock cycles.

\nocite{*}
\bibliographystyle{eptcs}
\bibliography{mybib}
\end{document}